\documentclass[doublecol]{epl2} 

\usepackage{amsmath,amssymb,enumerate,epsfig,bbm,calc,capt-of,ifthen}

\newcommand{\tmem}[1]{{\em #1\/}}
\newcommand{\tmmathbf}[1]{\ensuremath{\boldsymbol{#1}}}
\newcommand{\tmop}[1]{\ensuremath{\operatorname{#1}}}

\newcommand{\tmtextbf}[1]{{\bfseries{#1}}}
\newcommand{\tmtextit}[1]{{\itshape{#1}}}

\newenvironment{enumeratealpha}{\begin{enumerate}[a{\textup{)}}] }{\end{enumerate}}
\newenvironment{enumeratenumeric}{\begin{enumerate}[1.] }{\end{enumerate}}
\newenvironment{enumerateroman}{\begin{enumerate}[i.] }{\end{enumerate}}
\newenvironment{itemizedot}{\begin{itemize} }{\end{itemize}}
\newcommand{\tmfloatcontents}{}
\newlength{\tmfloatwidth}
\newcommand{\tmfloat}[5]{
  \renewcommand{\tmfloatcontents}{#4}
  \setlength{\tmfloatwidth}{\widthof{\tmfloatcontents}+1in}
  \ifthenelse{\equal{#2}{small}}
    {\ifthenelse{\lengthtest{\tmfloatwidth > \linewidth}}
      {\setlength{\tmfloatwidth}{\linewidth}}{}}
    {\setlength{\tmfloatwidth}{\linewidth}}
  \begin{minipage}[#1]{\tmfloatwidth}
    \begin{center}
      \tmfloatcontents
      \captionof{#3}{#5}
    \end{center}
  \end{minipage}}

\title{Thermodynamic versus Topological Phase Transitions: Cusp in the
Kert\'esz Line 
}

\author{Ph. Blanchard\inst{1} \and D. Gandolfo\inst{2} \and J. Ruiz\inst{2} \and M. Wouts\inst{3}}
\shortauthor{Ph. Blanchard \etal}

\institute{                    
  \inst{1} Fakult\"at f\"ur Physik, Theoretishe Physik and Bibos, Universit\"at
Bielefeld, Universit\"asstrasse, 25, D--33615, Bielefeld, Germany.  
\inst{2} Centre de Physique Th\'eorique, UMR 6207, Universit\'es Aix-Marseille et Sud Toulon-Var, Luminy Case 907, 13288 Marseille, France.
  \inst{3} Modal'X, Universit\'e Paris Ouest - Nanterre la D\'efense, 
B\^at. G,  200 avenue de la R\'epublique, 92001 Nanterre Cedex, France. 
}


\abstract{
  We present a study of phase transitions of the 
  mean--field 
  Potts model at
  (inverse) temperature $\beta$, in presence of an external field $h$. Both
  thermodynamic and topological aspects of these transitions are considered.
  For the first aspect we complement previous results and give an explicit
  equation of the thermodynamic transition line in the $\beta$--$h$ plane as
  well as the magnitude of the jump of the magnetization (for $q
  \geqslant 3)$. The signature of the latter aspect is characterized here by
  the presence or not of a giant component in the clusters of a
  Fortuin--Kasteleyn type representation of the model. We give the equation of
  the Kert\'esz line separating (in the $\beta$--$h$ plane) the two
  behaviours. As a result, we get that this line exhibits, as soon as $q
  \geqslant 3$, a very interesting cusp where it separates from the
  thermodynamic transition line.
}

\begin{document}

\maketitle


\section{
Introduction
}

In {\cite{Kertesz1}}, Kert\'esz pointed out that a very interesting phenomenon
arises in the Ising model subject to an external field. The so-called
Coniglio--Klein droplets {\cite{CK}} associated to Fortuin--Kasteleyn clusters
{\cite{FK1}} have a whole percolation transition line extending from the Curie
point  to infinite fields. This seems, at first sight in
contradiction with the fact that on the other hand, thermodynamics quantities
do not have any singularities in any of their derivatives (with respect to the
temperature or the field) as soon as the field is non--zero. But, as Kert\'esz
already mentioned:
\begin{quote}
  However, we emphasize that the suggested picture is not in contradiction
  with the non existence of singularities in the thermodynamic quantities
  because the total free energy remains analytic.
\end{quote}
Indeed in such a model, the criticality can be specified in two different
ways: the thermodynamic criticality associated to the thermodynamic limit of
the bulk free energy and the geometric criticality associated in the same
limit to another lower order free energy.

Since then, the Kert\'esz line has remained the subject of interests over the
years for various models of statistical mechanics. In particular, it has been
considered recently in {\cite{BGLRS}} within the Potts' model on the regular
lattice $\mathbb{Z}^d$. There it was found, that,  in dimension $d=2$, the whole thermodynamic
first order transition line (when such first order behavior is present in the
system) coincides with the Kert\'esz line. The latter is also first order in
the corresponding range of values of temperature and field.

The aim of this letter is to present a study of such a model when the
underlying lattice is the complete graph with $n$ vertices. 
In this case the
model is called mean--field or Curie--Weiss Potts model.

\section{The model}
To introduce the Potts model on the complete graph  
we attach  to the sites $i = 1, \ldots , n$ of the graph, spin variables $\sigma_i$ that take values in the set $\{1, \ldots, q\}$. 
The 
model at temperature $T = 1 / \beta$ and subject to an
external field $H = h / \beta$ is then defined by the Gibbs measure
\begin{equation}
  \mu_{\tmop{Potts}} (\tmmathbf{\sigma}) = \frac{1}{Z_{\tmop{Potts}}} 
  \prod_{i < j} e^{(\beta / n) (\delta_{\sigma_i, \sigma_j} - 1)} \prod_i e^{h
  \delta_{\sigma_i, 1}} \label{Potts}
\end{equation}
over spins configurations $\tmmathbf{\sigma}$. Here, $Z_{\tmop{Potts}}$
denotes the partition function, the indices $i, j$ runs over the set $\{1,
\ldots ,n\}$, and $\delta$ is the Kronecker symbol. The critical
(thermodynamic) behaviour of this model is well known {\cite{Wu}},
{\cite{CET}}, {\cite{BCC}}, and mainly governed by the mean field equation
\begin{equation}
  h = - \beta s + \ln \frac{1 + (q - 1) s}{1 - s} \label{MFE}
\end{equation}
Namely, when $h = 0$, there exists a threshold value
\begin{eqnarray}
  \beta_c^{(q)} & = & 2 \hspace{1em} \tmop{for} \hspace{1em} q = 2 
  \label{beta2}\\
  \beta_c^{(q)} & = & 2 \frac{q - 1}{q - 2} \ln (q - 1) \hspace{1em}
  \tmop{for} \hspace{1em} q \geqslant 3  \label{betaq}
\end{eqnarray}
such that the system exhibits at $\beta_c^{(q)}$ a continuous transition when
$q = 2$, and a first order transition when $q \geqslant 3$. When $h > 0$, no
transition occurs if $q = 2$, while as soon as $q \geqslant 3$ a first order
transition line appears for which 
the microcanonical free energy of the
model has
two minima associated with  
two different
solutions of the mean field equation   {\cite{Wu}}, {\cite{BCC}}.

To study the behaviour of clusters previously mentioned, we turn to the
Edwards--Sokal joint measure {\cite{ES}} given by 
\begin{multline}
  \mu_{\tmop{ES}} (\tmmathbf{\sigma}, \tmmathbf{\eta}) =
  \frac{1}{Z_{\tmop{ES}}}  \prod_{i < j} [e^{- \frac{\beta}{ n} } (1 - \eta_{ij}) + (1
  - e^{- \frac{\beta}{ n} }) \eta_{ij} \delta_{\sigma_i, \sigma_j}]
\\
\times 
\prod_i e^{h
  \delta_{\sigma_i, 1}} \label{ES}
\end{multline}
where the edges variables $\eta_{ij}$ belong to $\{0, 1\}$.

Notice that when $h = \infty$, all $\sigma_i = 1$. This means that we open
edges ($\eta_{ij} = 1$) with probability (w.p.) $p = 1 - e^{- \beta / n}$ and
close edges ($\eta_{ij} = 0$) w.p. $e^{- \beta / n}$. This is nothing else
than the well known Erd\"os--R\'enyi random graph $\mathcal{G}(n, p)$
{\cite{ER1}}. This random graph is known to exhibit a (topological) transition
at $\beta = 1$ such that with probability tending to $1$ as $n\to \infty$:
\begin{enumerateroman}
  \item for $\beta < 1$ all components of open edges are at most of order $\ln
  n$.
  
  \item at $\beta = 1$ a giant component of order $n^{2 / 3}$ appears.
  
  \item for $\beta > 1$ this giant component becomes of order $s^{\ast} n$
  where $s^{\ast}$ is the largest root of the mean field equation (\ref{MFE})
  with $q = 1$ and $h = 0$.
\end{enumerateroman}
We refer the reader to {\cite{Bo}}, {\cite{JLR}} for detailed discussions and
proofs (see also {\cite{BCS}} for a new approach of this transition).

Notice also that, on the other hand, when $h = 0$, the marginal of the ES
measure over the edges variables is the random cluster model:
\begin{equation}
  \mu_{\tmop{RC}} (\tmmathbf{\eta}) = \frac{1}{Z_{\tmop{RC}}}  \prod_{i < j}
  e^{- (\beta / n) (1 - \eta_{ij})} (1 - e^{- \beta / n})^{\eta_{ij}} q^{C
  (\tmmathbf{\eta})} \label{RC}
\end{equation}
where $C (\tmmathbf{\eta})$ denotes the number of connected components
(including isolated sites) of open edges of the configuration
$\tmmathbf{\eta}$. For $q = 1$, this model again reduces to $\mathcal{G}(n,
p)$ with $p$ as before. A refined study of the random cluster model (\ref{RC})
is given in {\cite{BGJ}} (see also {\cite{Grimmet}}, {\cite{LuLu}}). There, it
is shown that with the threshold value $\beta_c^{(q)}$, given by (\ref{beta2})
for $0 < q \leqslant 2$ and by (\ref{betaq}) for $q > 2$, then  
with probability tending to $1$ as $n\to \infty$:
\begin{enumeratealpha}
  \item if $\beta < \beta_c^{(q)}$, the largest component (of open edges) of
  $\mu_{\tmop{RC}}$ is of order $\ln n$.
  
  \item if $\beta > \beta_c^{(q)}$, the largest component of $\mu_{\tmop{RC}}$
  is of order $s_0 n$, where $s_0 > 0$ is the largest root of the mean field
  equation (\ref{MFE}) with $h = 0$.
  
  \item if $\beta = \beta_c^{(q)}$ and $0 < q \leqslant 2$, $\mu_{\tmop{RC}}$
  has largest component of order $n^{2 / 3}$.
  
  \item if $\beta = \beta_c^{(q)}$ and $q > 2$, $\mu_{\tmop{RC}}$ is either as
  in a) or as in b).
\end{enumeratealpha}

According to these results, it is natural to expect for model
$\mu_{\tmop{ES}}$ (and its marginal over the edges variables), a Kert\'esz
line $h_K (\beta)$, interpolating betwen $h_K(1) = \infty$ and $h_K
(\beta_c^{(q)}) = 0$, that signs the emergence of a giant component in the
open edges of the configuration $\tmmathbf{\eta}$.

To study this line, it is convenient to consider a colored version of the
Edwards--Sokal formulation. Namely, whenever the endpoints $ij$ of a given
edge are occupied by a spin of color $a$, we label the variable $\eta_{ij}$
with a superscript indicating the color. For a spin
configuration $\tmmathbf{\sigma}$, let $n_a (\tmmathbf{\sigma})$ be the number of
sites occupied by spins of color $a = 1, \ldots, q$.
We then relabel the variables $\eta_{ij}^{a}$ by 
$\eta_{kl}^{a}: k<l \in \{1,\dots,n_{a}(\tmmathbf{\sigma})\}$. 
This means that, for any
pairs $i < j  \in \{1,\dots,n_{a}(\tmmathbf{\sigma})\}$, we open an edge
$\eta_{ij}^a = 1$ w.p. $1 - e^{- \beta / n}$, and close this edge $\eta_{ij}^a
= 0$ w.p. $e^{- \beta / n}$. All the other edges between two sites of
different colors are closed. The resulting measure becomes
\begin{multline}
  \mu_{\tmop{CES}} (\tmmathbf{\sigma}, \tmmathbf{\eta}^1, \ldots,
  \tmmathbf{\eta}^q)  =  
\frac{e^{hn_1(\tmmathbf{\sigma})}}{Z_{\tmop{CES}}}  
\prod_{a = 1}^q e^{\beta n_a (\tmmathbf{\sigma}) (n_a
  (\tmmathbf{\sigma}) - 1) / 2 n} 
\\
   \times  \prod_{
i < j \in \{ 1, \ldots, n_a (\tmmathbf{\sigma}) \}
} 
[e^{- \beta /
  n} (1 - \eta^a_{ij}) + (1 - e^{- \beta / n}) \eta^a_{ij}]  
\label{CES}
\end{multline}
By summing over the spins variables, we get the following
Colored--Random--Cluster model
\begin{multline}
  \mu_{\tmop{CRC}} (\tmmathbf{\eta}^1, \ldots, \tmmathbf{\eta}^q)  = 
  \frac{1}{Z_{\tmop{CRC}}}  \sum_{n_1 + \cdots + n_q = n} Z (n_1, \ldots, n_q)
  \\
   \times  \prod_{a = 1}^q 
\prod_{i < j \in \{ 1, \ldots, n_a \}} 
[e^{- \beta x_a /
  n_a} (1 - \eta^a_{ij}) + (1 - e^{- \beta x_a / n_a}) \eta^a_{ij}] 
  \label{CRC}
\end{multline}
where $x_a = n_a / n$ are the densities of the colors and \
\begin{equation}
  Z (n_1, \ldots, n_q) = \frac{n!}{n_1 ! \cdots n_q !}\  e^{hn_1} \prod_{a =
  1}^q e^{\beta n_a (n_a - 1) / 2 n} \label{Pre}
\end{equation}
This colored random cluster model can be thought as follows. Given a partition
$n_1, \ldots, n_q$ of $n$, we have $q$ classical Erd\"os--R\'enyi random
graphs where the edges are open w.p. $1 - e^{- \beta x_a / n_a}$. The
asymptotic behavior of the partition as $n \rightarrow \infty$ will be
determined by minimizing the free energy
\begin{equation}
  \text{$f (x_1, \ldots, x_q) = - \lim_{n \rightarrow \infty} \frac{1}{n} \ln
  Z (n_1, \ldots, n_q)$} \label{fe}
\end{equation}
where the limit is taken in such a way that $n_a / n \rightarrow x_a$.

Notice that, up to a normalizing factor, $Z (n_1, \ldots, n_q)$ is the
microcanonical partition function of the Potts model (\ref{Potts}) restricted
to configurations such that the number of spins of color $a$ is fixed to
$n_a$. This will give the thermodynamic behavior of the system and will
determine the densities $x_a$. Once this is done the topological properties
can be analysed by using the known properties of classical random graphs. Due
to the presence of the field or using the symmetry of colors for vanishing
field, we will pick up the first color. Then, the threshold value of the
(topological) transition will be given by
\begin{equation}
  \beta_K x_1 = 1 \label{CT}
\end{equation}
For $\lambda = \beta x_1 > 1$ a giant component will appears, with size of
order 
$\Theta_{x_1}n\equiv \theta_{\beta x_1} x_1 n$ where $\theta_{\lambda}$ 
is the largest root of the mean field
equation
\begin{equation}
  \lambda \theta = - \ln (1 - \theta) \label{MFERC}
\end{equation}

Let us briefly recall the minimization procedure for (\ref{fe}). By  Stirling's
formula, we have $n! / n_1 ! \cdots n_q ! \sim \prod_{a = 1}^a e^{x_a \ln
x_a}$ at the leading order in $n$, giving
\begin{equation}
  f (x_1, \ldots, x_q) = \sum_{a = 1}^a x_a \ln x_a - \frac{\beta}{2}  \sum_{a
  = 1}^a x_a^2 - hx_1 \label{fe1}
\end{equation}
The minima of this function can be parametrised by a real number $s$, $0
\leqslant s \leqslant 1$, such that the minimizing vector $(x_1, \ldots, x_q)$
have the components
\begin{eqnarray}
  x_1 & = & \frac{1 + (q - 1) s}{q}  \label{x1}\\
  x_a & = & \frac{1 - s}{q} \hspace{1em} a = 2, \ldots, q  \label{xa}
\end{eqnarray}
(with arbitrary order for $h = 0$).

The free energy (\ref{fe1}) then takes the form
\begin{multline}
  f (s)  =  \frac{1 + (q - 1) s}{q} \ln ( 1 + (q - 1) s )
\\ +
  \frac{q - 1}{q} (1 - s) \ln (1 - s) 
   - \beta \frac{(q - 1)}{2 q} s^2 
\\ - h \frac{1 + (q - 1) s}{q}
 -
  \frac{\beta}{2 q} - \ln q  \label{fe2}
\end{multline}
The minimizers have to be found among the solutions of the mean field equation
(\ref{MFE}) obtained by differentiating $f$ with respect to $s$. Note that by
taking into account (\ref{x1}) the condition (\ref{CT}) becomes
\begin{equation}
  \beta_K  \frac{1 + (q - 1) s}{q} = 1 \label{CT2}
\end{equation}

\section{Results}
Let us first consider the case $q = 2$.

When $h = 0$, the mean field equation (\ref{MFE}) has a unique solution $s_0 =
0$ for $\beta \leqslant 2$, and a unique solution $s_0 > 0$ for $\beta > 2$.
Inserting $s_0$ into (\ref{CT2}), we obtain that the giant component appears
at $\beta_K^{(2)} = 2$, getting thus $\beta_K^{(2)} = \beta_c^{(2)}$. 
There, this component is of order $n^{2 / 3}$. For
$\beta > 2$, $\beta x_1 (s_0) > 1$, and we are in the supercritical regime
with a giant component of order $\Theta_{ x_1 (s_0)} n$. For $\beta < 2$, $\beta
x_1 (s_0) < 1$, and we are in the subcritical regime with  components of
order at most $\ln n$.

When $h > 0$, \ the mean field equation has a unique solution $s_0 > 0$ for
any $\beta$ (no thermodynamic transition occurs). The transition line
$\beta_K^{(2)} (h)$, or equivalently $h_K^{(2)}(\beta)$, is obtained by eliminating
$s$ between (\ref{MFE}) and (\ref{CT2}). This Kert\'esz line is given by
\begin{equation}
  h^{(2)}_K (\beta) = \beta - 2 - \ln (\beta - 1)
\end{equation}
On this line, the largest component is of order $n^{2 / 3}$. For $h < h_K$,
the largest component is of order $\ln n$, while for $h > h_K$, the largest
component is of order $\Theta_{ x_1 (s_0)} n$.

\begin{figure}[h]
 \begin{center}
   \epsfig{file=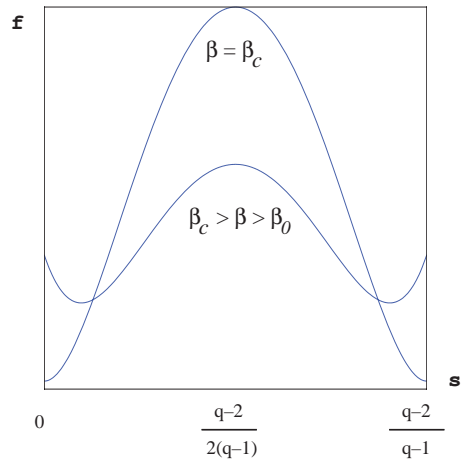, width=6cm}
  \caption{\label{Fs}The free energy on $h_T$.}
\end{center}
 \end{figure}

Let us 
then
consider the case $q \geqslant 3$. We have now a thermodynamic transition
line
\begin{equation}
  h_T (\beta) = - \beta \frac{q - 2}{2 (q - 1)} + \ln (q - 1), \hspace{1em}
  \beta_0 \leqslant \beta \leqslant \beta_c^{(q)} \label{hth}
\end{equation}
with endpoints
\begin{equation}
  (\beta_0 = 4 \frac{q - 1}{q}, h_0 =  - 2 \frac{q - 2}{q}+\ln (q - 1)),
  \hspace{1em} (\beta_c^{(q)}, 0) \label{EP}
\end{equation}
As previously mentionned, the thermodynamic transition has been already
established in {\cite{BCC}}, for a  curve with endpoints (\ref{EP}). 
To show that the thermodynamic curve is the straight line (\ref{hth})
we observe, on  $h_T$ 
the free energy (\ref{fe2}) has the symmetry $f (s) = f ( \frac{q
- 2}{q - 1} - s)$ with $s \in [0, \frac{q - 2}{q - 1}]$, see Fig. \ref{Fs}.
 
On $h_T$ the free energy has two minima $s_-$ and $s_+$ for which $f (s_-)
= f (s_+)$. They satisfy the mean field equation (\ref{MFE}) and thus are given by
the parametric equations
\begin{equation}
  (\beta, h) = (\beta_{s_{\pm}},  - \frac{q - 2}{2 (q - 1)}
  \beta_{s_{\pm}} +\ln (q - 1)) \label{s}
\end{equation}
where
\begin{equation}
    \beta_s  =  \left(s - \frac{q - 2}{2 (q - 1)}\right)^{- 1} \ln \frac{1 + (q - 1)
    s}{(q - 1) (1 - s)}
 \label{betas}
\end{equation}
and $s_- \in [0, \frac{q - 2}{2 (q - 1)}]$, $s_+ \in [ \frac{q - 2}{2 (q -
1)}, \frac{q - 2}{(q - 1)}]$.

\noindent For $\beta_0 < \beta \leqslant \beta_c^{(q)}$ (or equivalently $0 \leqslant h
< h_0$) these two minima are distinct: $s_- < s_+$. At ($\beta_0, h_0$), $s_-
= s_+ = \frac{q - 2}{2 (q - 1)}$, and at ($\beta_c^{(q)}, 0$), $s_- = 0$
and $s_+ = \frac{q - 2}{q - 1}$.

Outside of the segment (\ref{hth}), there is only one minimum $s_0$. This
minimum is an analytic function of $\beta \tmop{and}$ $h$, and $s_0 \rightarrow
s_-$ as $h \uparrow h_T$, $s_0 \rightarrow s_+$ as $h \downarrow h_T$.
In addition, on $h_T$, as $\beta$ increases from $\beta_0$ to $\beta_c^{(q)}$
(or equivalently as $h$ decreases from $h_0$ to $0$), 
$\beta x_1 (s_-)$
strictly decreases from 
$\beta_0 / 2 > 1$ 
to
$\beta_c^{(q)} / q < 1$, 
while $\beta x_1 (s_+)$ 
strictly increases from
$\beta_0 /2 $ 
to 
$\beta_c^{(q)}( q - 1)/q$.
See below.

We now turn to the topological behavior.

\noindent Let ($\beta_{\tmop{cp}}, h_{\tmop{cp}}$) the point of $h_T$ for which
$\beta x_1 (s_-) = 1$. This point is distinct from the two endpoints
(\ref{EP}). It is given by the solution of the equation $(\beta_{\tmop{cp}} -
2) q = 2 (q - 1) \ln (\beta_{\tmop{cp}}-1)$ which can be solved by using the
Lambert W function.

As a consequence of the above remarks, one gets the following topological
behaviour for the system. On the thermodynamic line $h_T$:
\begin{enumeratenumeric}
  \item at $\text{($\beta_{\tmop{cp}}, h_{\tmop{cp}}$)}$, the largest
  component of $\mu_{\tmop{CRC}}$ is either of order $n^{2 / 3}$ or of order
  $\Theta_{ x_1 (s_+)} n$.
  
  \item if $h_{\tmop{cp}} < h \leqslant h_0$, the largest component of
  $\mu_{\tmop{CRC}}$ is either of order $\Theta_{ x_1 (s_-)} n$ or of order
  $\Theta_{ x_1 (s_+)} n$.
  
  \item if $0 \leqslant h < h_{\tmop{cp}}$, the largest component of
  $\mu_{\tmop{CRC}}$ is either of order $\ln n$ or of order $\Theta_{ x_1 (s_+)}
  n$.
\end{enumeratenumeric}

The item 3 extends to the values $0 \leqslant h < h_{\tmop{cp}}$ what happens
in item d) of the random cluster model for vanishing field. It implies that
the thermodynamic and topological lines coincide there. This holds also at
$\text{$\text{($\beta_{\tmop{cp}}, h_{\tmop{cp}}$)}$}$, with a new behaviour
in the sense that the giant component is  of order $n^{2 / 3}$ or of
order $\Theta_{ x_1 (s_+)} n$.

\begin{figure}[h]
\begin{center}
  \epsfig{file=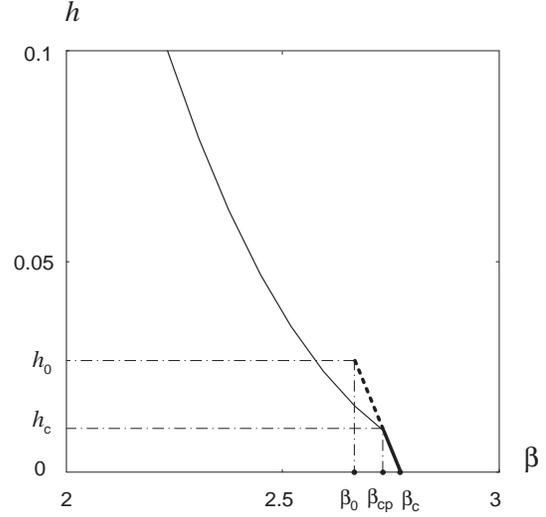,width=7cm,height=7cm}
  \vspace{0cm}
  \caption{\label{ttc}The thermodynamic transition curve:  thick (solid, dashed) line;
  the topological transition curve: solid (thick, thin) line.}
\end{center}
\end{figure}

The Kert\'esz line is given in this range of temperature (and field) by
\begin{equation}
  h_K^{(q)} (\beta) = h_T (\beta) \hspace{.8em} \tmop{for} \hspace{.8em}
  \beta_{\tmop{cp}} \leqslant \beta \leqslant \beta_c^{(q)} \hspace{.8em}
  \tmop{or} \hspace{.8em} 0 \leqslant h \leqslant h_{\tmop{cp}}
\end{equation}
The item 2 shows that the giant component exhibits a jump, however, for given
$h > h_{\tmop{cp}}$, this giant component already appeared for lower values of
$\beta$. The topological transition line is determined, as in case $q = 2$, by
eliminating $s$ between the mean field equation and the condition (\ref{CT2}).
This gives 
\begin{equation}
 h^{(q)}_K (\beta) = \frac{\beta - q}{q - 1} - \ln \frac{\beta - 1}{q - 1}
{\hspace{.8em}}\tmop{for} {\hspace{.8em}}
\beta \leqslant \beta_{\tmop{cp}} 
\hspace{.8em} \tmop{or} \hspace{.8em} 
h \geqslant h_{\tmop{cp}}
\end{equation}
see Fig.\ \ref{ttc}. Below the Kert\'esz line $h^{(q)}_K$ the largest component is of
order $\ln n$, while above $h^{(q)}_K$ and outside of the segment (\ref{hth}),
the largest component is of order $\Theta_{ x_1 (s_0)} n$.

The symmetry of the free energy is better seen if we parametrize the
magnetization $s$ by $z \in (\pm 1)$ as follows:
\begin{equation}
s = \frac{q - 2}{2 (q - 1)} + \frac{q}{2 (q - 1)} z. 
\end{equation}
The corresponding densities are given by
$
x_{1} = \frac{1 + z}{2}$,
$x_2 = \cdots = x_q = \frac{1 -
   z}{2 (q - 1)} 
$,
and the free energy decomposes immediately into its even and odd parts:
\begin{eqnarray}
  f (z) & = & \frac{1}{2}  [ (1 + z) \ln (1 + z) + (1 - z) \ln (1 - z) 
\nonumber\\
& &
- \beta \frac{q (1 + z^2)}{4 (q - 1)} - h - \ln \left( 4 (q - 1) \right)
  ] \nonumber\\
  &  & + \frac{z}{2}  \left[ \ln (q - 1) - \beta \frac{q - 2}{2 (q - 1)} - h
  \right] .  \label{foddeven}
\end{eqnarray}
The first derivative of $f$ yields the mean field equation (\ref{MFE}) while
the second derivative is
\begin{equation}
  f'' (z) = \frac{1}{1 - z^2} - \frac{\beta q}{4 \left( q - 1 \right)} .
\end{equation}

Thus f is convex when $\beta \leqslant \beta_0$ and has exactly one minima. When $\beta > \beta_0$ it has at most two local minimas. The analyticity of $s_0$ outside of $h_T$ is a consequence of $f''(z_0)>0$. The monotonicity of $\beta x_1 (s_{+})$ is trivial, that of $\beta x_1 (s_{-})$ can be infered from the computation of the contour lines for $\beta x_1(s)$ below $h_T$.

\section{Conclusion}
To summarize, we have given the equation of the Kert\'esz line of the mean--field Potts model
and shown that when $q\geqslant 3$ this line exhibits a cusp at some $(\beta_{\tmop{cp}},h_{\tmop{cp}})$.
For low fields ($h \leqslant h_{\tmop{cp}})$
the Kert\'esz line coincides with the thermodynamic transition line, and 
for large fields ($h \geqslant h_0)$ only the topological transition remains
while the thermodynamic transition disappears. In addition, at intermediate
fields ($h_{\tmop{cp}} < h < h_0)$ the Kert\'esz line separates from the
thermodynamic line. This means that decreasing the temperature one sees first
the appeareance of a giant component (on $h_K$) and, then on $h_T$, this
component exhibits a jump. This behavior is new compared to what happens for
the model on the $2$--dimensional regular lattice. There, no intermediate
regime appears {\cite{BGLRS}}. We expect for sufficiently high lattice
dimensions, a behaviour similar to the model on the complete graph.

\subsection{Acknowledgments}
Warm hospitality and Financial support from the
BiBoS Research Center, University of Bielefeld and Centre de Physique
Th\'eorique, CNRS Marseille are gratefully acknowledged.

\end{document}